\documentclass[11pt]{article}

\usepackage[letterpaper,top=2cm,bottom=2cm,left=3cm,right=3cm,marginparwidth=1.75cm]{geometry}

\usepackage{fancyhdr}

\usepackage{authblk}

\usepackage[numbers,sort&compress]{natbib}

\usepackage[colorlinks=true, allcolors=blue]{hyperref}
\usepackage{orcidlink}

\usepackage{amsmath}

\usepackage{graphicx}
\usepackage{subcaption}
\usepackage{placeins} 
\usepackage{xcolor}

\newcommand\blfootnote[1]{%
  \begingroup
  \renewcommand\thefootnote{}\footnote{#1}%
  \addtocounter{footnote}{-1}%
  \endgroup
}

\usepackage[
    type={CC},
    modifier={by-nc},
    version={4.0},
]{doclicense}
  
\title{3D Pore-Scale Mixing Interface Evolution}

\author[1]{Daniel M C Hallack\orcidlink{0000-0001-7171-0948}\thanks{Corresponding Author}}
\author[2]{Guillem Sole-Mari \orcidlink{0000-0002-9890-079X}}
\author[1]{Saif Farhat\orcidlink{0009-0000-3941-4049}}
\author[1]{Diogo Bolster\orcidlink{0000-0003-3960-4090}}

\affil[1]{{\scriptsize University of Notre Dame, Civil and Environmental Engineering and Earth Sciences}}
\affil[2]{{\scriptsize University of Rennes, Département Eau, Ressources et interactions fluides}}

\date{}

\begin{document}

{\let\newpage\relax\maketitle}

\hrule
\begin{abstract}
The effective mixing behavior of solutes in porous media is fundamentally connected to the development of a local mixing interface between the two initial solutions, which is characterized by a complex lamellar structure. The deformation of the interface is controlled by the interplay of advection and diffusion, which generate the mechanisms of lamella  stretching and shrinking, respectively. Based on the results of pore-scale numerical simulations, we develop a mechanistic single parabolic lamella model (SPLM) to capture the interface evolution across various temporal and P\'eclet number scales. The model shows near-perfect agreement with a 2D parallel plates scenario and promising results for a 3D porous medium. The SPLM model also establishes P\'eclet regimes for the equilibrium area and temporal regimes for the transient behavior of the interface. These findings represent a step forward towards eventually incorporating mixing limitation into general macroscopic reactive transport models.


\end{abstract}
\vskip  8 pt
\hrule

\blfootnote{E-mail addresses:dhallack@nd.edu}



\vspace{-20 pt}
\section{Introduction}
\label{sec:intro}

Reactive transport limited by mixing in porous media is a widespread phenomenon in the natural environment (\cite{dentz2011mixing, valocchi2019mixing}), spanning many sectors of interest, such as groundwater contamination and remediation (\cite{sebilo2013long,dybas2002development}), CO2 sequestration and enhanced oil recovery (\cite{szulczewski2012lifetime, niemi2017geological}, \cite{jimenez2016mixing}) and stream substrate ecology (\cite{aubeneau2015effects, angang}) to name a few. Mixing serves as the essential process by which reactants are brought together, allowing for their subsequent reaction (\cite{valocchi2019mixing}). Therefore, the development of models capable of predicting mixing processes is essential as a starting point for a reaction model(e.g. \cite{de2005procedure}). Within the context of porous media, the large number of distinct spatial and temporal scales that can be involved and compete with one another makes upscaling by conventional means difficult, if not impossible (e.g. \cite{battiato2009breakdown}, \cite{battiato2011applicability}). The complex structure of a porous medium leads to velocity variations at the pore scale that are not resolved at the Darcy scale, which in turn has heterogeneities that are not often resolved at the field scale. This lack of representation at larger scales introduces incomplete mixing behaviors, a phenomenon that has been observed in laboratory experiments multiple times (\cite{gramling2002reactive, raje2000experimental, oates2007upscaling}) as well as in the field (\cite{ding2017elimination}). Not accounting for these can lead to significant mismatches between observations and predictions. In this study we will focus on mixing at pore scales and specifically in the context of a three-dimensional porous medium.

To understand and predict mixing, a useful approach is to study interfaces which potential reactants must cross for reactions to happen. In two dimensions such interfaces are lines, whereas in three-dimensions they are surfaces. The interface's shape and temporal evolution reflect the flow dynamics of the system and are intricately related to the mixing process (\cite{de2014filamentary,valocchi2019mixing}). In the case of a continuous injection of a conservative solute, one may track the isocontour corresponding to the midpoint concentration between the injected and displaced solution concentrations, which represents a critical interface that would physically divide reactants in the context of instantaneous reactions.  In principle, any concentration isocontour or similar interface could be tracked, but consistent behaviors are expected (\cite{farhat2024evolution}). 

At the pore scale, this interface deforms through two primary mechanisms: advection and molecular diffusion. The former involves a non-uniform velocity distribution due to the complex makeup of a porous medium and its interaction with a viscous fluid flowing through it. These non-uniform velocities lead to the spreading of solutes, leading to the stretching of the surface and the formation of filamentary structures, often referred to as lamellae (\cite{le2015lamellar,le2013stretching,villermaux2012mixing}), or diffusive strips (\cite{meunier2010diffusive}). By contrast, molecular diffusion acts to homogenize solute concentration fields, reducing the elongation of filaments and thus decreasing the net deformation of the surface. A simple way to quantify this behavior in an aggregate manner is to examine the evolution of the total interfacial surface area over time. 

 To date, Lagrangian models of solute transport have been employed to investigate the interaction of these two mechanisms and their influence on mixing interfaces, predominantly in 2D systems. \citet{le2013stretching} use the lamellae concept at the Darcy scale to analytically predict an upscaled concentration PDF for a given  P\'eclet number and medium heterogeneity. \citet{de2014filamentary} describe the dynamics of the lamellae at the pore scale, identifying two temporal regimes. The early-time stretching regime is dominated by flow kinematics, whereas the late-time coalescence regime is dominated by a random aggregation process. 
 Though limited, there has also been some recent research about lamellae in 3D systems. \citet{lester2022fluid} describes Darcy-scale lamellae stretching in the framework of a continuous time random walk(CTRW) in three dimensions.
 Pore-scale velocity fields in 3D have an additional degree of freedom, allowing more intricate and even chaotic behaviors (\cite{olaf}, \cite{wunsch2000numerical}).
 Chaotic mixing models have been proposed to explain some of the scalings observed in pore-scale 3D experimental observations (\cite{sanquer2024microscale,heyman2020stretching}).

Advanced high-performance computing and high-resolution simulations at the pore scale have made it possible to look much more closely at the evolution of concentrations (\cite{icardi2014pore},\cite{branko}) and thus also advance the investigation of mixing and reaction processes in porous media \cite{branko2}. Most such simulations feature very small porous media specimens due to computational constraints, and while they provide valuable insight, they may miss important larger-scale behaviors. Recently, the simulations carried out by \citet{sole2022closer} reproduced a full-scale mixing-front laboratory experiment (such as, e.g., \cite{gramling2002reactive}).
These simulations offer an unparalleled opportunity to explore details unattainable in typical laboratory experiments and are ideally suited to study mixing and concentration interface dynamics.

In this work, based on the evolution and asymptotic behaviors of the interface area extracted from the \citet{sole2022closer} numerical simulations, we identify the main scaling features and propose a model to help understand the physics behind them. The model accounts for the effects of both advection and diffusion on the evolution of a mixing lamellar interface across a wide range of temporal and P\'eclet number scales.  

This paper is structured as follows. In \autoref{sec:data} we describe the extraction of the concentration isosurfaces from the simulation data. We present the surface area evolution data obtained and identify the main scaling features. In \autoref{sec:SPLM} we propose a parsimonuous single parabolic lamella model (SPLM) to partially capture the observed behaviors. In \autoref{sec:discussion} we evaluate the performance of the model for two scenarios, flow between two parallel plates and the 3D porous medium, discussing strengths and limitations. Finally, the summary and conclusions are presented in \autoref{sec:conclusion}.  

\section{Pore-Scale Data}
\label{sec:data}

\subsection{Simulation description}
\begin{figure}[htp]
  \centering
    \includegraphics[width=0.9\textwidth]{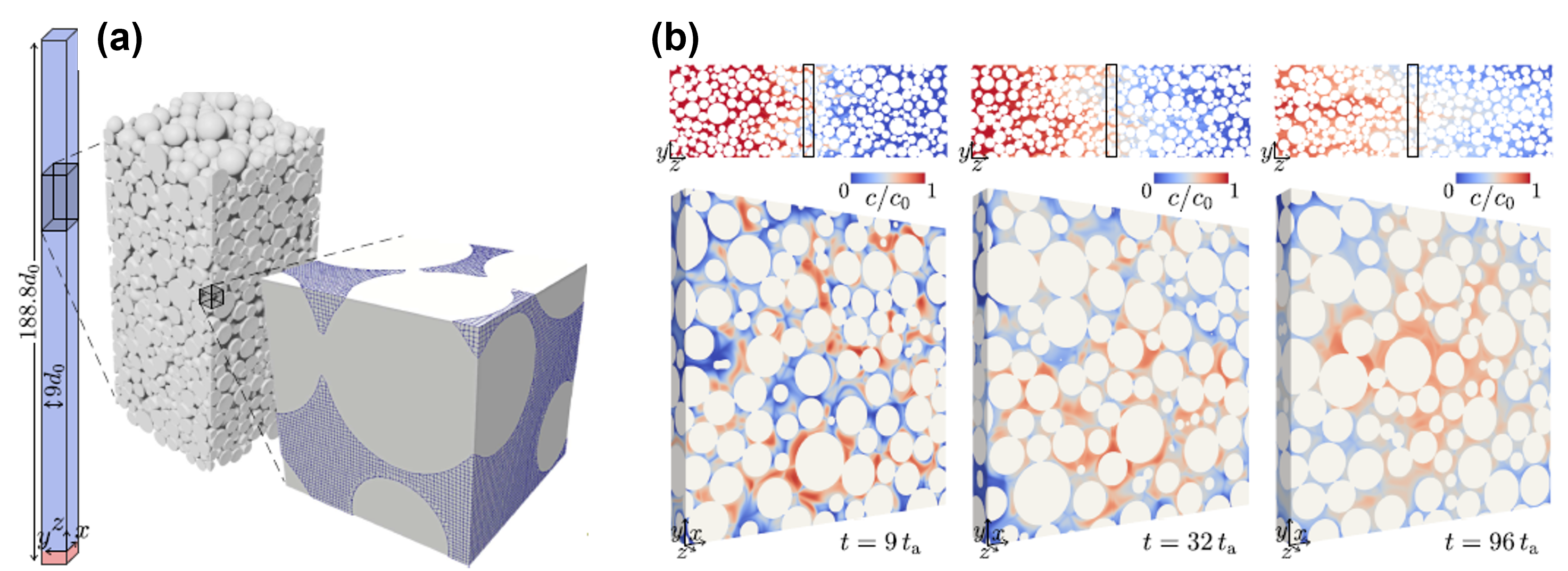}
    \caption{(a) Full simulation column dimensions (9$d_0$, 9$d_0$, 188.8$d_0$) , an example detailed portion of it, and mesh used for numerical simulation. (b) Simulation results shown at three different times on a slice-cut following the moving mixing front.}
    \label{fig:Column}
\end{figure}

All work presented here is based on solute transport data obtained by the simulations of \citet{sole2022closer}. We provide a brief description here but direct the interested reader to that paper for full details. In order to gain a better understanding of experimental observations of mixing such as those of \citet{gramling2002reactive}, \citet{sole2022closer} built a digital random porous medium made of spherical solid grains of uniform diameter $d_0$, resembling a sand-filled column, and simulated flow and transport using the OpenFOAM suite \cite{OpenFOAM}. The medium was created by letting grains settle by gravity using the software Blender \cite{blender}. The interstitial space was meshed with a cubic regular mesh with a cell size of $d_0/60$ which was then transformed into an unstructured one to capture grain-fluid interface geometries accurately. \autoref{fig:Column}(a) shows the full column dimensions, an example portion of it, and a finite volume mesh. This mesh was then used to solve the steady-state, incompressible Navier-Stokes equations,

\begin{equation}
\label{1}
(\Vec{u}\cdot\nabla)\Vec{u}= \nu\nabla^2\Vec{u}-\frac{1}{\rho}\nabla p,
\end{equation}
to obtain the velocity field at the pore scale. Here, {$\Vec{u}$} denotes the velocity vector, {$\rho$} is the fluid density, {$\nu$} is the kinematic viscosity, and {$\nabla p$} is the pressure gradient. No-slip boundary conditions are imposed at the fluid-grain boundaries, while a full-slip boundary condition is imposed on the column lateral boundaries so as to minimize their influence on flow and ultimately transport. The advection-diffusion equation was then solved to simulate the transport of a solute,
\begin{equation}
    \label{2}
    \frac{\partial C} {\partial t} = -\Vec{u}\cdot\nabla C+D\nabla ^2 C,
\end{equation}
where the solute concentration is denoted by {$C$}, and the diffusion coefficient by {$D$}. 

The function $C(t=0)=1-H(x-x_0)$ was used as the initial condition for the transport simulation to mimic an initial sharp flat interface between two solutions near the inlet, where $H(x)$ is the Heaviside step function. No-flux boundary conditions were imposed at the fluid-grain boundaries and at the column lateral walls. The simulation was run for a sufficiently long time for the invading fluid to arrive at the column exit, whereupon it is terminated. Six P\'eclet number cases ($Pe={10,32,100,316,1000,3160}$) were simulated by modifying the molecular diffusion coefficient, where $Pe=\overline{u}d_0/D$, where $\overline{u}$ is the average fluid velocity and $D$ the diffusion coefficient.
 A full description of the numerical simulation setup is given in \cite{sole2022closer}.

\subsection{Generation of mixing interfaces}

\begin{figure}[htp]
  \centering
    \includegraphics[width=0.9\textwidth]{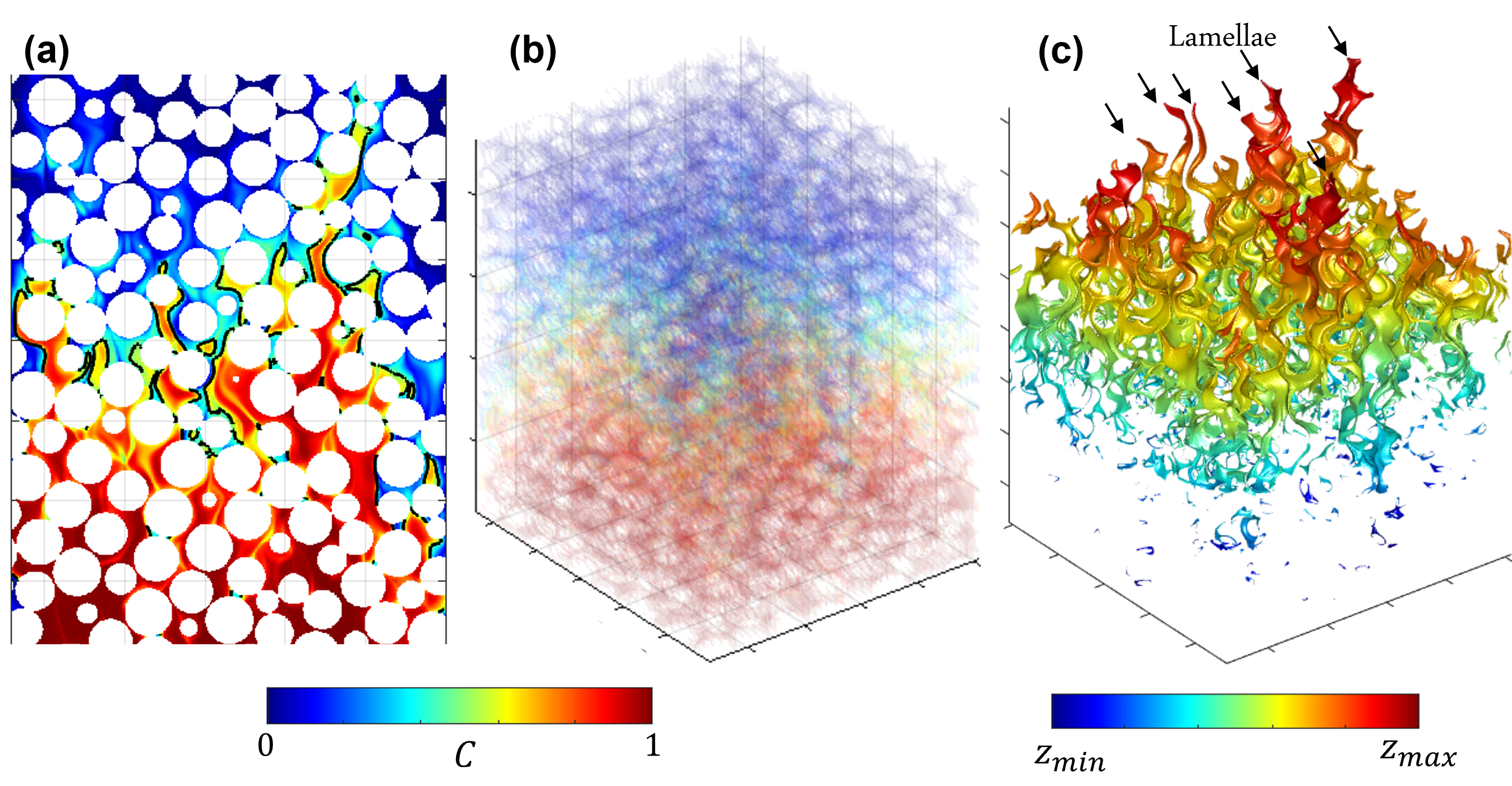}
    \caption{(a) 2D cross-section of a concentration field (left colorbar legend) and the 0.5 contour line (black line) describing the interface; (b) 3D concentration field (left colorbar legend); (c) 3D 0.5 concentration isosurface (right colorbar legend, colors represent vertical distance, not concentration). Elongated structures along the flow direction are referred to as "lamellae".}
    \label{fig:iso}
\end{figure}

The interface between two solutions at the pore scale is not expected to be sharply defined as diffusion leads to a transition zone rather than an abrupt boundary. The concentration gradually shifts from $1$ to $0$. We identify the mixing interface as the iso-concentration contour for the midpoint (i.e., $0.5$) concentration value. A representative example of such an isosurface is shown in \autoref{fig:iso}. 

The 3D isosurfaces are generated by interpolating the concentration field using Matlab's \textit{isosurface} function which is based on the Marching Cubes algorithm (\cite{lorensen1987marching}). These surfaces are described by a fine triangular mesh that is able to resolve structures at smaller scales than the pore size (\autoref{fig:iso}(c)).

\begin{figure}[htp]
  \centering
    \includegraphics[width=1\textwidth]{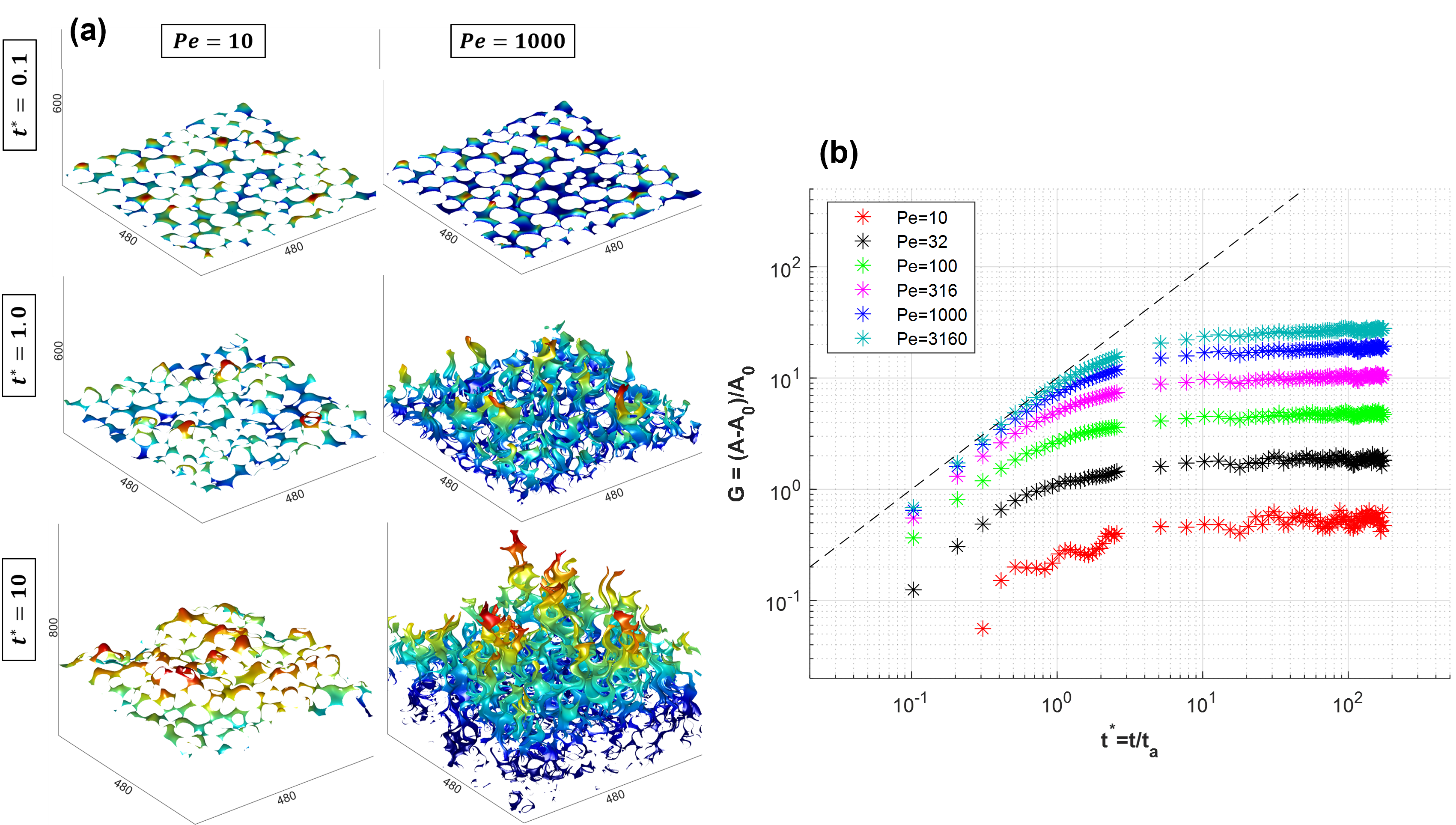}
    \caption{Evolution with time of (a) the mixing interfaces and (b) their area growth for different Péclet numbers.}
    \label{fig:isoPe}
\end{figure}

The complex filamentary structures that arise from this type of analysis (\autoref{fig:iso}(c)) are typically referred to as lamellae (\cite{le2015lamellar},\cite{villermaux2012mixing}), and will be called as such from now on in this paper. Their overall geometry at different times for P\'eclet numbers 10 and 1000 is depicted in \autoref{fig:isoPe}(a). As expected, for the higher P\'eclet number, advective stretching is stronger relative to diffusion and this leads to more elongated lamellae.

We quantify the deformation of the iso-concentration interface by the growth of its total surface area, 
\begin{equation}
    \label{3}
   G(t)  = \frac{A(t)-A_0}{A_0},
\end{equation}
where $A_0$ is the initial flat interface area at time $t=0$. 

By computing this quantity over time and across the range of P\'eclet simulations available, we obtain the evolution profiles presented in \autoref{fig:isoPe}(b), where time is normalized by the advection time, defined as $t_a = \frac{d_0}{\bar u}$. The main features that stand out from \autoref{fig:isoPe} can be summarized as

\begin{enumerate}
  \item The interface area grows with time but eventually approaches a stagnation plateau.
  
  \item At early times, the interface area growth exhibits a faster than linear behavior. 
  
  \item The plateau, i.e., the late time stagnation area value, increases with P\'eclet number, but not linearly. 
\end{enumerate}

In order to understand this behavior, rather than focusing directly on the system at hand, in the following section we will begin by exploring the mechanisms at play in the idealized context of a single parabolic fluid velocity profile.

\section{Single Parabolic Lamella Model} \label{sec:SPLM}

As a starting point to modeling the observed interface behavior presented in the previous section, we hypothesize that the evolution of a single lamellar structure could be sufficiently representative of the ensemble of lamellae that make up the surface (\autoref{fig:iso}(c)). Such ideas have been useful in other upscaling contexts in porous media (e.g. \cite{juanes1,lazaro}). 

\begin{figure}[htp]
  \centering
    \includegraphics[width=1\textwidth]{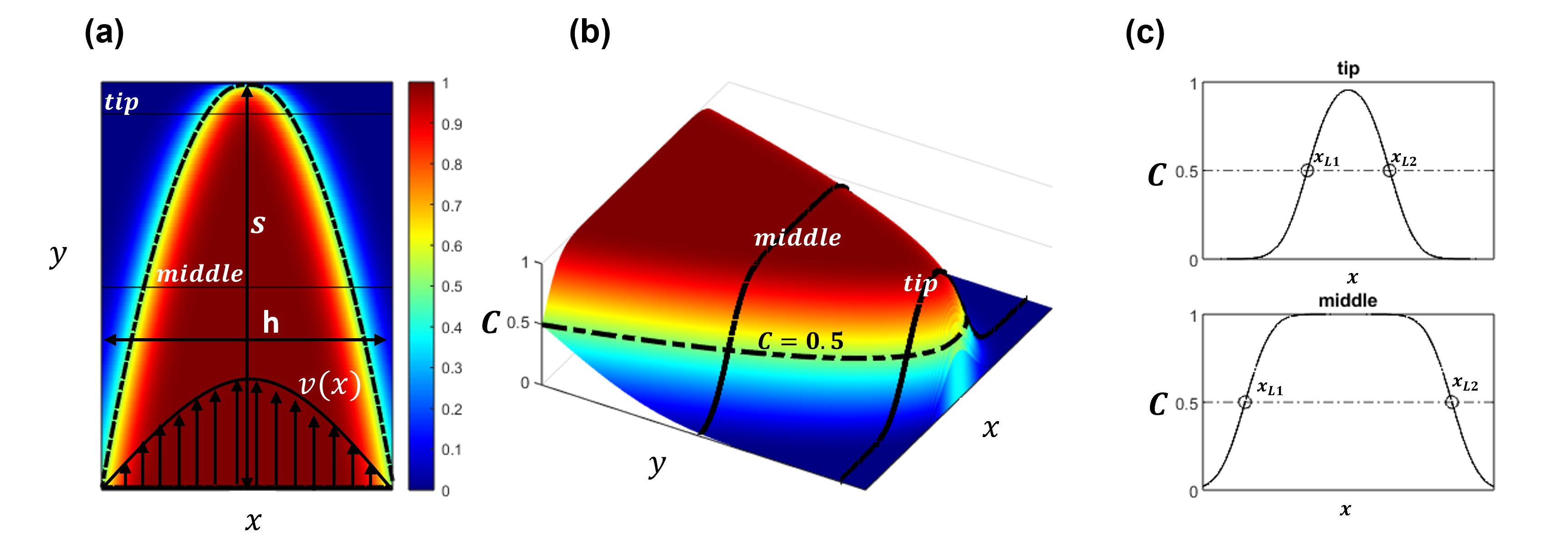}
    \caption{(a) Top view and (b) 3D (xyC) view of 2D concentration field representing a parabolic lamella. The 0.5 concentration isoline is the mixing interface line for the single parabolic lamella model (SPLM). (c) Concentration profiles perpendicular to flow direction at the tip and middle of the lamella.}
    \label{fig:2Dlamella}
\end{figure}

To begin, consider a 2D parabolic lamella subjected to a parabolic velocity field as depicted in \autoref{fig:2Dlamella}. 
The contour geometry is defined by the length $s$ and width $h$ (\autoref{lam_profile}). The velocity field $v(x)$ is defined by the average velocity $\bar{v}$ and width $h$ (\autoref{vel_profile}) and assumed to not depend on flow direction $y$: 

\begin{equation}
    \label{lam_profile}
    y_L(x)  = 4s \frac{x}{h} \bigg( 1 - \frac{x}{h} \bigg),
\end{equation}

\begin{equation}
    \label{vel_profile}
    v(x)  = 6 \bar v \frac{x}{h} \bigg( 1 - \frac{x}{h} \bigg).
\end{equation}

As mentioned previously, two primary competing mechanisms shape the evolution of the lamella length ($s$): advection and molecular diffusion. If we assume that these can be decoupled from one other, we might say that they respectively generate a stretching rate ($\frac{\partial s}{\partial t}\big|_{stretch}>0$) and a shrinking rate ($\frac{\partial s}{\partial t}\big|_{shrink}<0$). 

\subsection{Stretching Rate}

In \autoref{fig:2Dlamella}(a), a lamella is stretched by the velocity profile $v(x)$, since the middle travels at a higher velocity than the borders. The total stretch rate can be obtained by integrating the velocity gradient $\partial v/\partial x$ over the half-width of the lamella,  

\begin{equation}
    \label{5}
    \frac{\partial s}{\partial t}\bigg|_{stretch}  = \int_0^{h/2} \frac{\partial v(x)}{\partial x} dx = c_A \bar v,
\end{equation}
where $c_A$ is a dimensionless stretching proportionality constant, representing the geometry and spread of a velocity profile acting on the lamella contour. For the two-dimensional lamella representation proposed above, it is $3/2$. This changes, for instance, with dimension and boundary geometry, such that, for a 3D lamella with a circular paraboloidal profile, we have $c_A=2$. One should therefore select the suitable value of $c_A$ depending on the dimensionality and geometry of the system at hand. 
Aside from that, the stretching rate is only dependent on the average flow velocity.

In a porous medium, we may assume that lamellae are generated by a similar stretching mechanism to the above, as near-parabolic velocity profiles occur between nearby solid walls, albeit with the added distortion and heterogeneity brought about by the random geometry \cite{juanes1}.
 However, the parallel plates model's assumption of constant velocity along streamlines is a crucially unrealistic representation of an actual porous medium. Within the latter, velocity correlation along the flow direction can be statistically characterized by a typical distance $\lambda$. A hypothetical long lamella such that $s\gg\lambda$ would tend to experience a significantly diminished stretching rate. Indeed, given two points of the same lamella separated by a sufficiently large longitudinal distance, their two-point advective stretching rate (their mutual separation velocity) would appear random with near-zero mean due to decorrelation.

To account for this, we introduce a penalty to the stretching rate, which is determined by the ratio of lamella length ($s$) to velocity correlation length ($\lambda$), such that

\begin{equation}
    \label{stretch_corr}
    \frac{\partial s}{\partial t}\bigg|_{stretch}  = \frac{c_A  \bar v }{1 + \frac{s}{\lambda}}.
\end{equation}

When lamellae are short ($s<\lambda$) their limited spatial extent experiences a highly correlated velocity profile, resulting in no significant penalty to the stretching rate. By contrast, for longer lamellae ($s>\lambda$) the penalty becomes substantial as the less-correlated velocity field loses its stretching ability.
The exact format of the proposed correction of the stretching rate (\autoref{stretch_corr}) will be discussed and justified based on the observed data.

\subsection{Shrinking Rate}

 Associated with the idealized parabolic lamella profile, we can construct an also idealized 2D concentration field $C$, defined by \autoref{C_field}, which is a superposition of two complementary error functions ($\textrm{erfc()}$) perpendicular to the lamella direction (\autoref{fig:2Dlamella}(c)). Such functions were chosen because it is a typical solution for the diffusion equation. They are centered at the lamella contour line $x_L$, which is the inverse function of the proposed parabolic profile \autoref{lam_profile} ($x_L(y) = y_L^{-1}(x)$). 
The parameter $Dt_0$ sets a non-zero initial horizontal spread to create a continuous concentration field (\autoref{fig:2Dlamella}) that smoothes out with time.

\begin{equation}
    \label{C_field}
    C(x,y,\Delta t) = \frac{1}{2} \textrm{erfc}\left(\frac{x - h\frac{1+\sqrt{1-y/s}}{2}}{2\sqrt{D(t_0+\Delta t)}}\right)-
    \frac{1}{2} \textrm{erfc}\left(\frac{x - h\frac{1-\sqrt{1-y/s}}{2}}{2\sqrt{D(t_0+\Delta t)}}\right). \\
\end{equation}

\begin{figure}[htp]
  \centering
    \includegraphics[width=0.8\paperwidth]{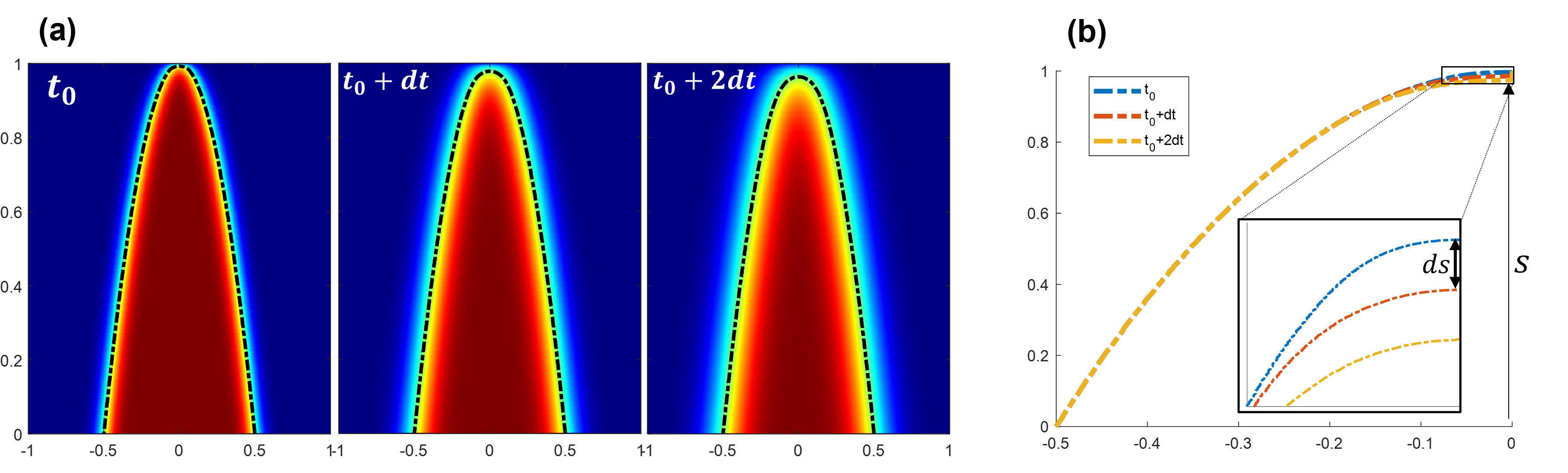}
    \caption{(a) Concentration field of a lamella evolving over small time increments under diffusion, with a dashed line representing the lamella contour; (b) Lamella length $s$ shrinking by diffusion, with zoom at the tip.}
    \label{fig:shr}
\end{figure}

Molecular diffusion ($D$) has an impact on the lamella contour line by changing concentration field $C$. We assume that only transverse diffusion contributes significantly to this effect, and this assumption is corroborated by numerical observations in a parallel plate setup. 
We use \autoref{C_field} to evaluate the change in the lamella contour ($C(x,y_L,t)=0.5$) after a small change in time $dt$. \autoref{fig:shr} shows how diffusion changes the lamella contour line at its tip after one and two time steps ($dt$).

By numerically evaluating the change in $s$ after a time increment $dt$ under diffusion, and how this rate is affected by geometric parameters $s$ and $h$, we find the following differential equation:

\begin{equation}
    \label{shrink}
    \frac{\partial s}{\partial t}\bigg|_{shrink}  = - c_D D s\left(\frac{1}{h}\right)^2,
\end{equation}
 where $c_D$ is a dimensionless shrinking proportionality constant. Since this mechanism generates a negative change in the length, we call it shrinking rate. This scaling was found by dimensional analysis, and in fact, this general diffusive shrinking behavior, other than the particular value $c_D$, does not depend on the parabolic shape assumption or lateral boundary conditions. More details on how \autoref{shrink} was obtained can be found in \autoref{app_A}. 
 
\subsection{Equilibrium condition and transient behavior}

From the previously derived relationships we may write a differential equation for the lamella length:

 \begin{equation}
    \label{dstotal}
    \frac{ds}{dt}  = \frac{\partial s}{\partial t}\bigg|_{stretch} + \, \frac{\partial s}{\partial t}\bigg|_{shrink} = \frac{c_A  \bar v }{1 + \frac{s}{\lambda}} - \frac{c_D D s}{h^2}.
\end{equation}
 
The equilibrium state ($\frac{ds}{dt}=0$) is attained when stretching and shrinking rates become equal in magnitude, resulting in a stabilization of both lamella length and contour length, which corresponds to our plateau regime. This equilibrium involves a balance between the mechanisms of stretching driven by heterogeneous advection and shrinking driven by molecular diffusion.
 
Defining Peclet number as $Pe = \frac{\bar v h}{D}$, $s^*=s/h$, $\lambda^*=\lambda/h$, $c = \frac{c_A}{c_D}$, and setting \autoref{dstotal} to zero, we obtain a dimensionless lamella length at equilibrium as

 \begin{equation}
    \label{sequil}
    {s}^*_{equil}  = \frac{\lambda^*}{2} \left( \sqrt{1 + 4 \frac{cPe}{\lambda^*}}-1\right).
\end{equation}

From the role of $Pe$ in the above equation, we note that the higher the advection forces in comparison to diffusion forces, the longer a lamella can get at equililbrium, aligning with intuitive expectations. 
Additionally, we can identify two limiting cases:
 \begin{equation}
    \label{sequil_lowPe}
    {s}^*_{equil}  \approx c Pe, \ \mathrm{for} \ cPe\ll\lambda^*,
\end{equation}
 \begin{equation}
    \label{sequil_highPe}
    {s}^*_{equil}  \approx \sqrt{c Pe\lambda^*}, \ \mathrm{for} \ cPe\gg\lambda^*.
\end{equation}

\noindent It is worth noting that \autoref{sequil_lowPe} also corresponds to the limiting case of infinite longitudinal velocity correlation length ($\lambda^*\rightarrow\infty$), such as in an idealized infinite tube.

 When the two rates are not equal, we are in a transient regime of the lamella evolution. At early times when $s^*=0$, the lamella is expected to grow (\autoref{fig:isoPe}) because the stretching rate overcomes the shrinking rate. Defining non-dimensional time $t^*=t\,\bar v/h$ we may rewrite \autoref{dstotal} as:
 
\begin{equation}
    \label{netrate}
   \frac{\partial s^*}{\partial t^*} =   \frac{c_A}{1+\frac{s^*}{\lambda ^*}}  - c_D \frac{s^*}{Pe},
\end{equation}
which can be integrated to obtain the transient $s^*(t)$ for any $Pe$. While we cannot obtain a general analytical solution, we can do so for specific limiting cases (similar to the two equilibrium limiting cases shown above):

\begin{equation}
    \label{sshort}
    s^*(t^*)  \approx cPe(1-e^{- \frac{c_D}{Pe}t^{*}}), \ \mathrm{for} \ s^*\ll\lambda^*,
\end{equation}
\begin{equation}
    \label{slong}
    s^*(t^*)  \approx \sqrt{c Pe \lambda^*(1-e^{-2 \frac{c_D}{Pe}t^{*}})}, \ \mathrm{for} \ s^*\gg\lambda^*.
\end{equation}

In the first limit (\autoref{sshort}), lamellae are shorter than the velocity correlation length, which occurs at early times and for lower Peclet numbers. Conversely, the second limit (\autoref{slong}) characterizes the evolution of elongated lamellas, observed at later times and for higher Peclet numbers. The complete solution (obtained numerically) offers a more comprehensive description of the transition between these two regimes.  

\subsection{Length and Area growth} 

As noted in Section \autoref{sec:data}, the measured mixing interface is an area in a 3D medium and a length in a 2D medium. In order to compare such measurements to our model predictions, one needs to convert lamella length, $s$, to either interface length, $L$, or interface area, $A$.

Based on the parabolic shape described by \autoref{lam_profile}, in two dimensions we have:

\begin{equation}
    \label{length_growth}
    G  = \frac{L-h}{h} =
    \frac{1}{h} \int_0^h\sqrt{1+\left(\frac{\partial y_L}{\partial x}\right)^2}dx - 1 =
    \frac{asinh(4s^*)+4s^*\sqrt{(4s^*)^2+1}}{8s^*}-1,
\end{equation}
where $G$ is the relative interface length growth.
By exploiting radial symmetry, the two-dimensional (2D) parabolic lamella can be converted into a three-dimensional (3D) paraboloid. In this case we have 

\begin{equation}
    \label{area_growth}
    G = \frac{A-A_0}{A_0} =
    \frac{1}{A_0} \int_0^{h/2}\sqrt{1+\left(\frac{\partial y_L}{\partial x}\right)^2}2\pi xdx - 1 =
    \frac{((4s^*)^2+1)^{3/2}-1}{24s^*{}^2}-1,
\end{equation}
where $G$ becomes the relative interface area growth, and $A_0=\frac{\pi h^2}{4}$ is the flat initial interface area.

From both equations \eqref{length_growth} or \eqref{area_growth}, two scaling limits can be identified by using the leading terms of series expansions: when the lamella is much shorter than its width  ($s^* \ll 1$) the interface grows quadratically with $s^*$, and when the lamella is much longer than its width ($s^* \gg 1$) the interface grows linearly with $s^*$.

By combining \autoref{netrate} with \autoref{length_growth} or \autoref{area_growth}, we can estimate interface growth as a function of time ($G(t)$) for any given $Pe$. By combining \autoref{sequil} with \autoref{length_growth} or \autoref{area_growth}, we can estimate the equilibrium interface area as a function of $Pe$, $G_{equil}(Pe)$.

 \subsection{Scaling regimes} \label{scalings}

\begin{figure}[htp]
  \centering
    \includegraphics[width=0.7\paperwidth]{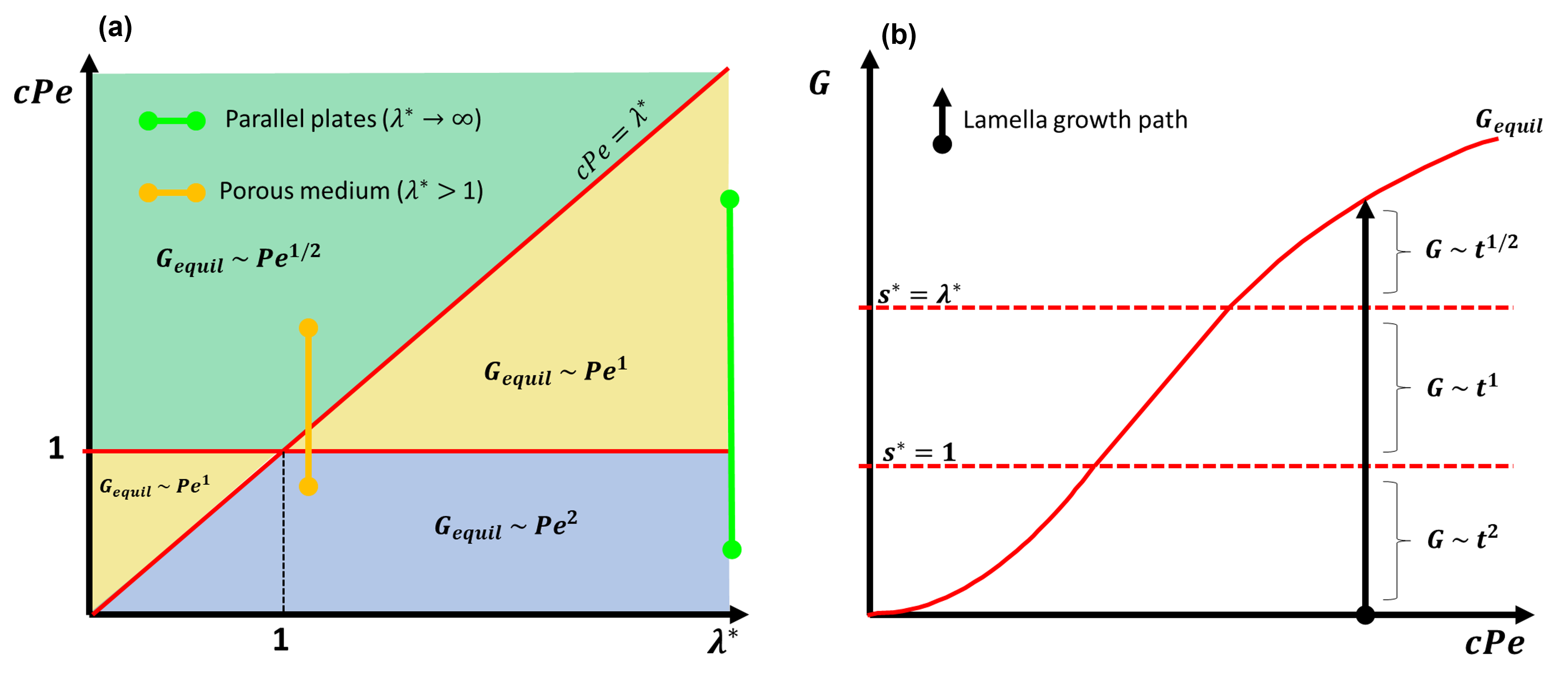}
    \caption{(a) Zones of equilibrium growth scaling with P\'eclet. "Parallel plates" and "Porous medium" are the cases analyzed in \autoref{sec:discussion};  (b) Transient growth scaling with time, illustrated for some arbitrary finite $\lambda^*>1$. The vertical arrow represents  the path, in terms of temporal regimes, that a growing lamella follows until reaching equilibrium for a given P\'eclet number.}
    \label{fig:diagrams}
\end{figure}

Based on the SPLM model described, we can anticipate the expected scaling regimes for the equilibrium interface area with P\'eclet (\autoref{fig:diagrams}(a)). And in the same way, predict regimes for the evolution of the interface with time (\autoref{fig:diagrams}(b)). Below we describe the expected regimes for $\lambda^*>1$, as is the case for our porous medium data presented in \autoref{sec:data}. The regimes for the equilibrium interface area are:    
 
\begin{enumerate}
    \item For low $Pe$ cases, characterized by $cPe \ll 1$, $G_{equil} \sim Pe^2$.
    
    \item For intermediate $Pe$ cases, characterized by $ 1 \ll cPe \ll \lambda^*$, $G_{equil} \sim Pe^1$.
    
    \item For high $Pe$ cases, 
    characterized by $ cPe \gg \lambda^*$, $G_{equil} \sim Pe^{1/2}$.
    
\end{enumerate}

\noindent For the interface area evolution with time, the following regimes are identified:
 
 \begin{enumerate}
 \item At very early times, characterized by $s^* \ll 1$, $G \sim t^2$.

 \item During intermediate times, when $1 \ll s^* \ll \lambda^*$, $G \sim t$.

 \item At later times, with $s^* \gg \lambda^*$, $G \sim t^{1/2}$.

 \item The interface ultimately stagnates, reaching a plateau regime ($s^*=s^*_{equil}, G = G_{equil}$).
 
 \end{enumerate}

 However, not all these regimes may manifest clearly under specific conditions. For instance, if the equilibrium is reached in earlier stages ($s^*_{equil}<\lambda^*$ or $s^*_{equil}<1$), as shown in the parallel plates case in \autoref{fig:diagrams}(a) or if velocity correlation length is not much larger than the width ($\lambda^* \sim 1$), as shown in the porous medium case in \autoref{fig:diagrams}(a). Likewise, \autoref{fig:diagrams}(b) illustrates the different stages of lamella growth, which follows a vertical line from $G=0$ to $G=G_{equil}$. Depending on the value of $\lambda^*$ and on that of $G_{equil}$ (which depends on $cPe$ and $\lambda^*$), some of the temporal growth regimes may not occur.

\section{Results and Discussions} \label{sec:discussion}

Our single parabolic model (SPLM) is first validated against an idealized 2D parallel plates flow case and then compared against the porous medium dataset presented in \autoref{sec:data}.

\subsection{2D parallel plates} \label{sec:2Dsection}

\begin{figure}[htp]
  \centering
    \includegraphics[width=1\textwidth]{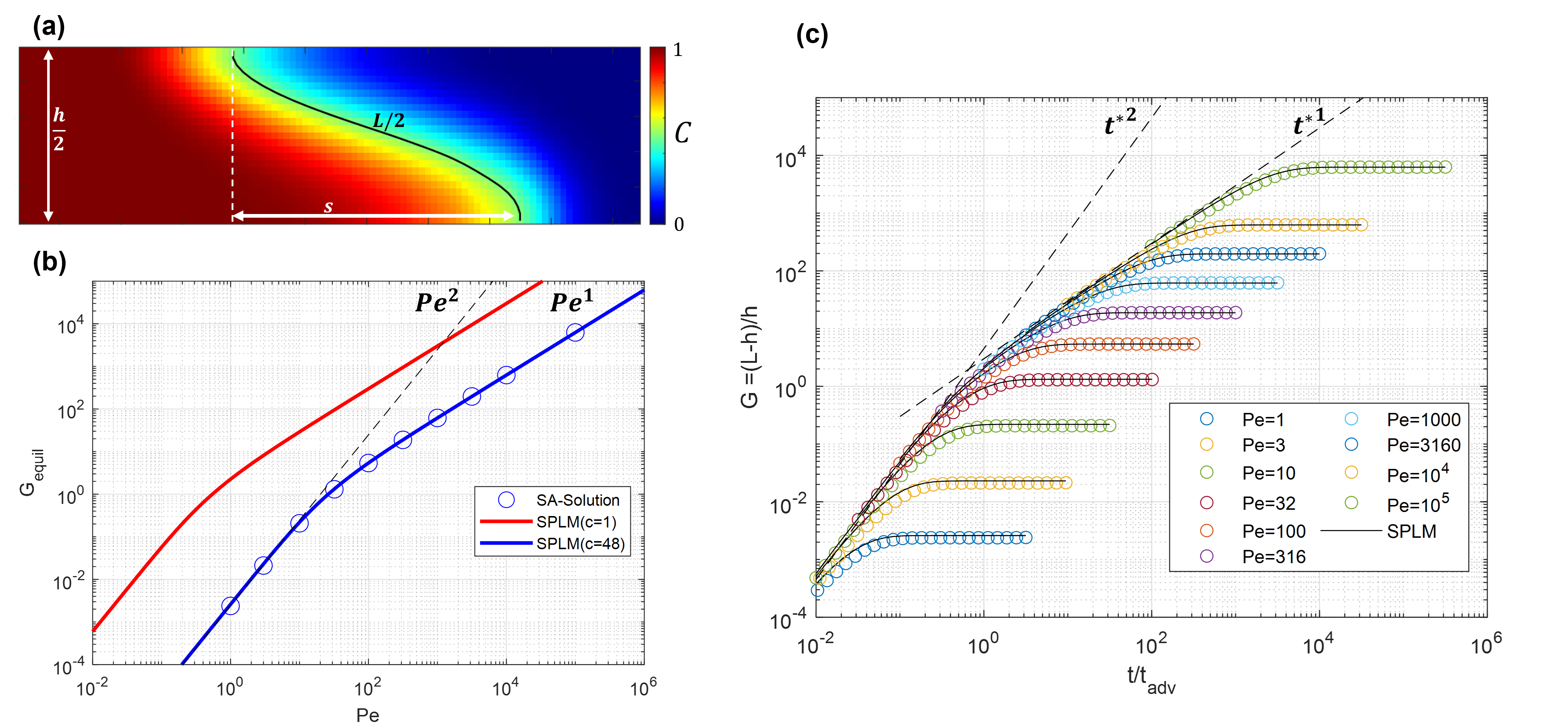}
    \caption{2D parallel plates case. (a) Concentration field and mixing interface contour snapshot; (b) Relationship between $Pe$ and equilibrium interface growth (SPLM model matches semi-analytical solution for $c=1/32$) (c) Transient growth behavior for all $Pe$, with SPLM model comparison ($c=1/32$).}
    \label{fig:2Dcase}
\end{figure}

Let us consider a continuous tracer injection into a fixed pressure-driven flow between two parallel plates (2D Poiseuille Flow). The mixing interface (0.5 concentration isoline) develops into a single lamella, as shown in \autoref{fig:2Dcase}(a), which matches the idealized model, yet without the parabolic shape. This setup is a traditional benchmark problem (\cite{juanes1, lazaro}) which has both numerical and semi-analytical solutions that enable accurate calculations of interface size evolution over time and at equilibrium.

In this configuration, the constant velocities along streamlines imply that we are in the limit case of $\lambda^* \rightarrow \infty$. 
Therefore, the model for the equilibrium interface length is defined by Equations \ref{sequil_lowPe} and \ref {length_growth}, and the model for its temporal evolution is defined by Equations \ref{sshort} and \ref {length_growth}. From the Stokes flow solution of the velocity profile, we have $c_A = 3/2$, thus the sole parameter in the model that remains to be determined is the proportionality constants ratio  $c=\frac{c_A}{c_D}$.  

We compare the solution of the SPLM with a semi-analytical solution \cite{farhat2024evolution} for a wide range of $Pe$, from $1$ to $10^5$. Using the equilibrium interface length ($G_{equil}$) relation with P\'eclet , we determine the proportionality constants ratio value $c=1/32$ to adjust the model to the reference solution \autoref{fig:2Dcase}(b). 

The entire transient regime of the interface growth behaves as predicted, with a near-perfect agreement between the model and the semi-analytical solution \autoref{fig:2Dcase}(c) for the entire range of $Pe$ tested.
 
Both the equilibrium and transient behaviors (Figures \ref{fig:2Dcase}(b) and \ref{fig:2Dcase}(c), respectively) show scaling regimes as described in subsection \ref{scalings} and \autoref{fig:diagrams}, with the exception (in both cases) of regime 3, which cannot be reached since $\lambda^*\rightarrow\infty$. For low $Pe$ cases ($Pe<32$), the transient regime 2 in which $G\sim t$ is skipped because $s^*_{equil} < 1$ and therefore the condition $1 \ll s^* \ll \lambda^*$ never occurs. 

It has been verified (not shown here for brevity) that the same type of results and model matching are obtained from a three-dimensional parallel plates (square pipe) setup, after adjusting $c$ adequately, confirming that the applicability of the model does not depend on the system's dimensions.


\subsection{3D Porous medium}

In this scenario, we use the numerical simulation data described in section \autoref{sec:data}. The SPLM model is applied, using the grain size diameter $d_{0}$ as the representative single lamella width $h$, and assuming a circular Poiseuille flow, which gives paraboloid velocity profile with $c_A = 2$.  
Dealing in this case with a 3D problem, we compute the interface area growth by \autoref{area_growth} to compare it to the data.  

For the velocity correlation length $\lambda$, we estimate it based on the velocity variogram along streamlines, which has been found to be about one grain diameter  ($1.2 \pm 0.6 d_0$). This represents the typical distance along a streamline over which velocity retains a certain degree of autocorrelation, and such an estimate is in line with our expectations that it should be on the same order of magnitude as the grain diameter (e.g., \cite{capuani2003correlation}). 

\begin{figure}[h!]
  \centering
    \includegraphics[width=1.0\textwidth]{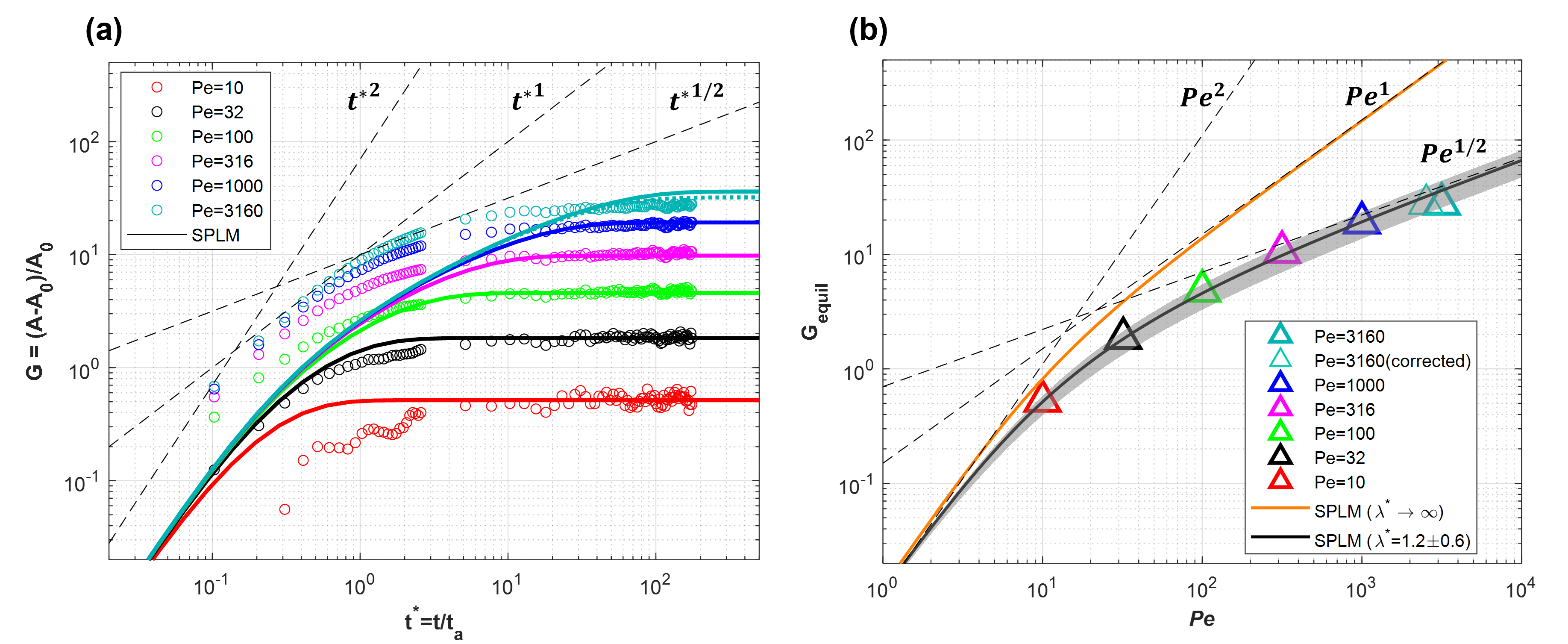}
    \caption{3D porous medium case. Interface growth predicted by the model(SPLM) versus simulation data for (a) Transient behavior, and (b) Equilibrium. Solid lines represent the model and markers represent the data. For $Pe$ 3160, a minimum estimated impact of numerical dispersion ($D_{num} \approx 25\% D_m$) is shown by a left shift on the respective $G_{equil}$ marker. For other $Pe$ numbers, this impact was negligible. The grey area band represents the uncertainty in the model due to the uncertainty in the velocity correlation length, $\lambda^*$.}
    \label{fig:PMcase}
\end{figure}

\autoref{fig:PMcase} shows the comparison between SPLM and the porous medium measurements for several values of $Pe$. Due to small fluctuations in the data at late times, we use the average of the last 10 time points as representative of the equilibrium interface area for all cases. 
In the same way that we did for the parallel plates configuration, choosing an adequate $c$ to fit the equilibrium data ($c=1/18$), the model captures reasonably well the behavior seen in the data (\autoref{fig:PMcase}(b)). Although significant efforts were undertaken to minimize such effects, since the data is based on a numerical simulation, it can be impacted by numerical dispersion, particularly for our largest Peclet number. An assessment under idealized uniform flow conditions, has allowed us to estimate numerical dispersion for the largest P\'eclet case to be at least $D_{num} \approx 25\% D_m$. This artifact may be interpreted as an effectively simulated P\'eclet number which is lower than the intended value. This is illustrated in  \autoref{fig:PMcase}(b) by a leftward shift of the respective marker and in \autoref{fig:PMcase}(a) by the dashed line representing the SPLM model for a P\'eclet number after correction. For other P\'eclet numbers ($Pe\leq10^3$), this impact was negligible. 

The difference between the lines SPLM($\lambda \rightarrow \infty$) and SPLM($\lambda=1.2$) shows the impact of velocity correlation length and how it is crucial to capture the data. With regards to the transient behavior our model captures the general trend, but broadly does not reflect the early time behaviors as well as one would hope, missing the spread observed among different $Pe$ at very early times (\autoref{fig:PMcase}(a)). 
We suspect that there is a sizable impact of pore-space heterogeneity in the early-time advective stretching, such that in the actual porous medium a variety of lamellae are locally generated at various rates and with uneven geometry. In contrast, the model is based on single effective values of lamella width $h$ and velocity $\bar v$, thus disregarding heterogeneity, which explains the much lesser impact that $Pe$ has on early-time interface growth compared to the data. This effect becomes less important at later times, when the various lamellae have a statistically similar history and can thus be modeled by a single set of effective parameters.
The fact that the plateau is reached in all cases within a few advective times suggests that transient effects wash out fairly quickly and are only important within a few grains of the initial sharp interface. 

Finally, we can back up our proposed stretching rate formulation (\autoref{stretch_corr}) based on the observed porous media data scalings in \autoref{fig:PMcase}(b). We note that $G_{equil} \sim \sqrt{Pe}$ for large $Pe$, which means $s_{equil} \sim \sqrt{Pe}$ for $s^* \gg 1$. Given that $\frac{\partial s}{\partial t}|_{shrink} \sim s^{1}$ (\autoref{shrink}), it must be that $\frac{\partial s}{\partial t}|_{stretch} \sim s^{-1}$ for sufficiently large $Pe$, such that at equilibrium $s_{equil} \sim \sqrt{Pe}$ be satisfied. As expected, the larger the value of P\'eclet, the longer the lamellae become and a more significant limitation is imposed by velocity decorrelation effects (separation between orange and black lines in \autoref{fig:PMcase}(b)).   

Regarding observed regimes (see Section 3.5 and \autoref{fig:diagrams}), it is worth noting that regime 2 for the equilibrium ($G_{equil} \sim Pe^1$)  is skipped or very short. This is due to the fact that the velocity correlation length is very similar to the grain size, or the lamella width, i.e., $\lambda^* \sim 1$, hence the condition $ 1 \ll cPe\ll \lambda^*$ is very short-lived. The same idea can explain why regime 2 ($G \sim t$ for $1\ll s^*\ll \lambda^*$) does not appear clearly in the transient behavior. Regime 1, both equilibrium ($G_{equil} \sim Pe^2$) and transient ($G \sim t^2$), are also difficult to observe in the data due to the numerical limitations for very low $Pe$ and for very short times, respectively.

\subsection{Early times behavior remarks}
As noted, the early time predictions from our current model do not capture the spread of behaviors with $Pe$ seen in the numerical data, which we primarily attribute to a heterogeneous distribution of parameters in the actual system as opposed to the simple model.  To this end we delve a little deeper to better try and capture early time effects. Applying series expansions on the transient lamella length equations \ref{sshort} and \ref{slong}, we notice that the stretching proportionality constant $c_A$ controls the lamella growth at early times in our model:

\begin{equation}
    \label{lim_sshort}
    \lim_{t^*\to 0} s^*(t)  \approx c_A t^* , \ \mathrm{for} \ s^*\ll\lambda^*,
\end{equation}

\begin{equation}
    \label{lim_slong}
    \lim_{t^*\to 0} s^*(t)  \approx \sqrt{2c_A \lambda^*t^*} , \ \mathrm{for} \ s^*\gg\lambda^*.
\end{equation}

\begin{figure}[h!]
  \centering
    \includegraphics[width=0.95\textwidth]{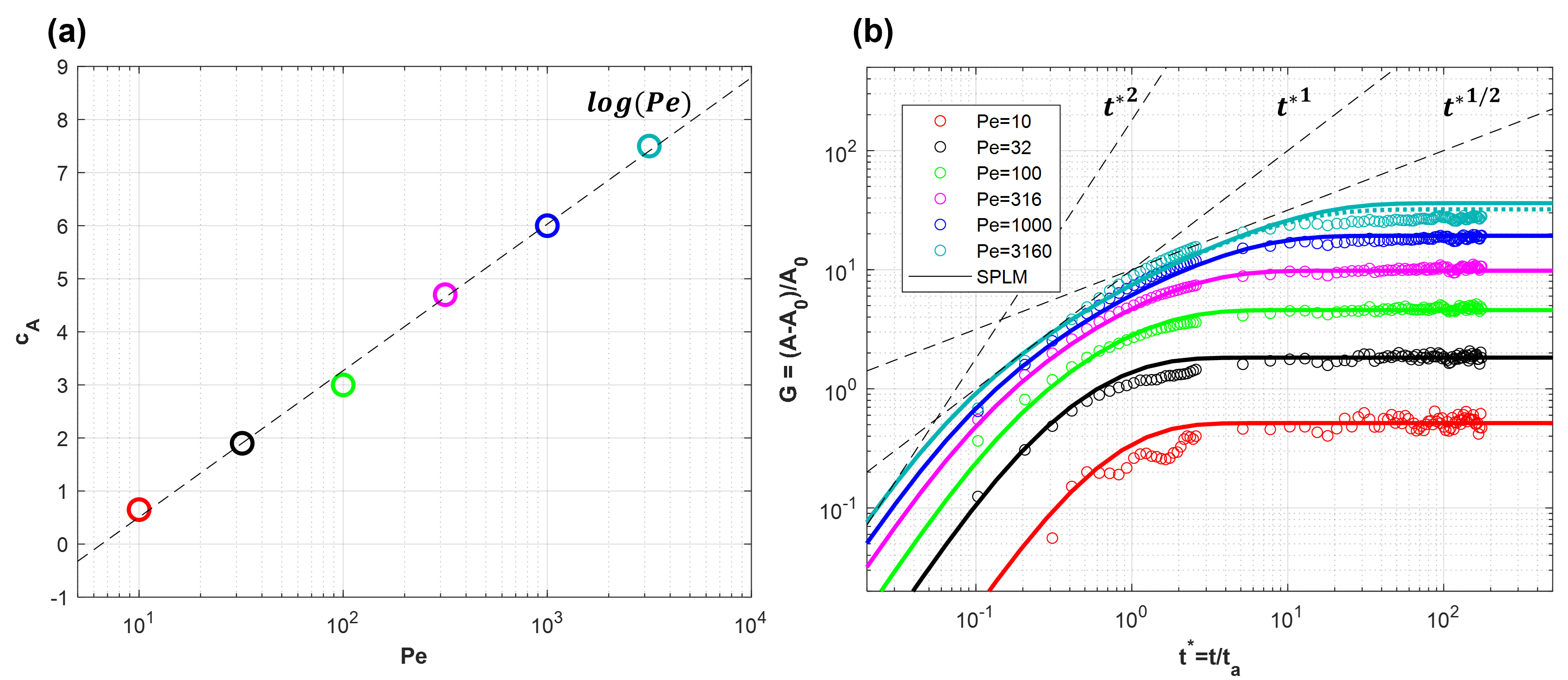}
    \caption{3D porous medium case. (a) The relationship observed after fitting of $c_A(Pe)$. (b) Interface growth is predicted by the model using variable $c_A$ versus simulation data for transient behavior.}
    \label{fig:fitcv}
\end{figure}

The spread seen at early times in the porous medium data across different P\'eclet numbers (\autoref{fig:PMcase}(a)) suggests a dependency of $c_A$ on P\'eclet. Executing a fitting procedure on $c_A$ we observe $c_A \sim log(Pe)$ (\autoref{fig:fitcv}(a)). Note that to keep the equilibrium solution, which matches the data this would mean $c_D$ must have the same scaling with P\'eclet ($c_D \sim log(Pe)$).


Assuming this relationship comes from the fact that we are modeling an ensemble of lamellae instead of a single one, we can think of this as a filter applied on the original model (\autoref{netrate}) that arises due to averaging to represent the porous medium behavior also at early times. In other words 

\begin{equation}
    \label{dstotal_ensemble}
    \bigg\langle \frac{ds}{dt} \bigg\rangle  = \left( \frac{c_A}{1+\frac{s^*}{\lambda ^*}}  - c_D \frac{s^*}{Pe} \right) \bigg(\alpha+\beta log(Pe)\bigg).
\end{equation}

where $<>$ denotes the averaging operator. When accounting for this a reasonable match for all times, including early ones, is obtained as shown in \autoref{fig:fitcv}(b). 

Despite our attempts, at this stage we do not have a mathematical demonstration of how this factor would arise. However, we do note that similar scalings depending on $log(Pe)$ arise naturally in classical studies of upscaling dispersion in porous media such as that off Saffman (\cite{saffman1959theory}) when deriving residence times in a pore-network approximation of an ensemble of particles. We believe that there may be an analogy here.  

\section{Conclusion} \label{sec:conclusion}

Based on observations from a series of column-scale numerical simulations with pore-scale resolution, we have proposed a simple model for the evolution and equilibrium of the mixing interface (midpoint iso-concentration surface) between two solutions. The model is built up from a single lamella with a fixed parabolic geometry under the influence of two competing physical mechanisms: advection (stretching) and diffusion (shrinking). A single fitting parameter $c$, whose value is likely medium-dependent, provides a scale to the strength ratio of lamellae stretching and shrinking.

The model is initially conceived based on a fixed velocity profile (Poiseuille flow) and then developed to account for velocity fluctuations along streamlines, which is characteristic of flows in porous media. In the lower P\'eclet number ($Pe$) regime, diffusion can kill lamella growth before it reaches a length comparable to the velocity correlation length.  In the larger $Pe$ regime, where longer lamellae are developed, then growth is influenced and limited by the longitudinal decorrelation of velocities. This notion is somewhat analogous to theories for the asymptotic dispersion coefficient transitioning from Taylor to Scheidigger type dispersion. 

Our relatively simple SPLM model is able to capture the following observed features in the data: (1) the interface size always tends to an equilibrium; (2) interface growth at early times is faster than linear and (3) the equilibrium size grows nonlinearly with P\'eclet. The model shows near perfect agreement with results from a 2D parallel plates case and also promising agreement (considering the model's simplicity) with results from a more complex 3D porous medium. In this case, the model excels particularly at capturing the equilibrium state across a comprehensive range of Péclet numbers. 

The model can be adapted to agree with the early-time data in the porous medium if a filter depending on $log(Pe)$ is considered. Our interpretation is that the model's depiction of the system by one single lamella and one characteristic length might not be sufficient to characterize the heterogeneous ensemble, while at later times, given sufficient sampling of the velocity field, it is.

The interface growth at equilibrium ($G_{equil}$) is described analytically as a function of $Pe$, given the velocity field's correlation length ($\lambda$). The model identifies (and the data corroborates) the existence of up to 3 scaling regimes, where the equilibrium growth $G_{equil}$ scales as $Pe^2$, $Pe^1$ or $Pe^{1/2}$, depending on which range of $Pe$ the system is in. During transient behavior (before equilibrium is reached), the interface area growth is described as a function of time, given a system's P\'eclet number and velocity correlation length. We identify up to 4 possible temporal regimes for the interface growth temporal scaling ($t^2$, $t^1$, $t^{1/2}$ and equilibrium).

One interesting case that we were unable to explore in our setups would be a highly anisotropic system where the velocity correlation length is much higher than the characteristic width/grain size ($\lambda^*\gg1$). In such conditions, it should be possible to see all the expected regimes for the equilibrium interface area. This may be a subject of future work.

\section*{Acknowledgements}

This research was funded by NSF Grant EAR2049688 and by the European Commission (MixUp, MSCA-101068306).

\section*{Data availability}
The data for this work are available at GitHub: \url{https://github.com/dhallackla/isosurface_data}.


\appendix
\section{Shrinking model}
\label{app_A}

\autoref{fig:shr} shows how diffusion alters the lamella contour line. The majority of the contour line remains fixed/close to its initial location (\autoref{fig:shr} (b)) because changes in the concentration gradient slope do not impact the interface's position. However, as seen in the blown-up part of the figure, the mutual merging of the concentration fields from both sides results in the interface's backwards migration at the center, leading to the collapse of the tip and a reduction in lamella length. 

To evaluate the new lamella length after a change in time $dt$ we solve the following equation for $y=s_{new}$ at the center $x=h/2$:
\small
\begin{equation}
\label{length}
C\bigg(x=\frac{h}{2},y=s_{new},t=t_0+dt\bigg) = \frac{1}{2}\left[ \textrm{erfc}\left(-\frac{h}{4}\sqrt{\frac{1-s_{new}/s}{D(t_0+dt)}}\right)-\textrm{erfc}\left(\frac{h}{4}\sqrt{\frac{1-s_{new}/s}{D(t_0+dt)}}\right)\right]  =  0.5.
\end{equation}
\normalsize

By solving for $s_{new}$ we calculate the shrinking rate. Although we cannot find a direct, explicit equation for it, by inspection we can see that the shrinking rate can only depend on four parameters: width ($h$), length ($s$), molecular diffusion ($D$), and initial time ($t_0$) which sets the initial gradient, such that

\begin{figure}[htp]
\centering
\includegraphics[width=0.6\textwidth]{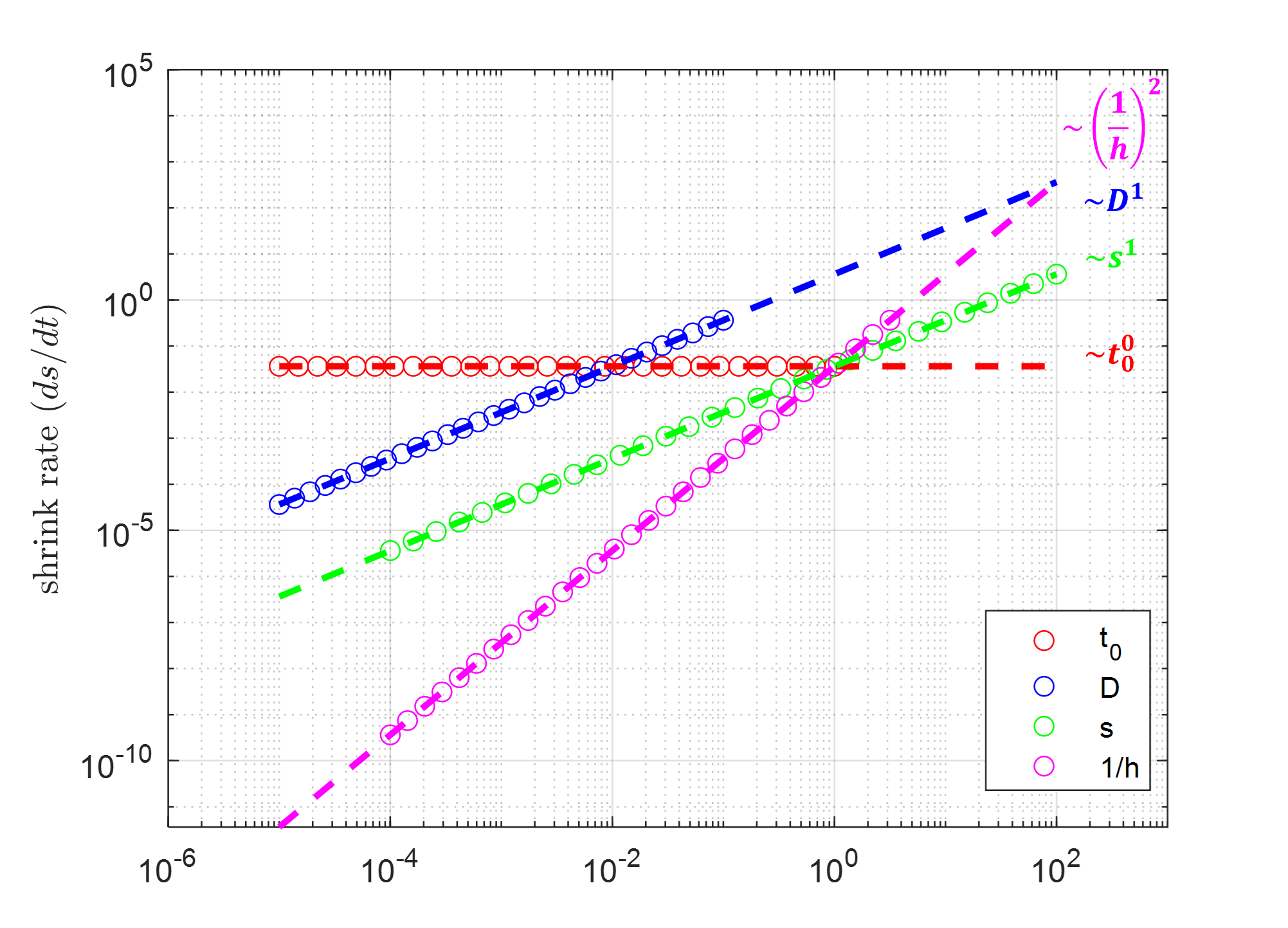} 
    \caption{Shrinking rate scalings with diffusion (D),  length (s), width (h) and initial gradient ($t_0$).}
    \label{fig:shrrate}
\end{figure}

\begin{equation}
\label{shrinkrate1} 
 \frac{\partial s}{\partial t}\bigg|_{shrink}  =
 \frac{s-s_{new}}{dt}  = f(h,s,D,t_0).
\end{equation}

The influence of each of these parameters on the shrinking rate can be empirically estimated via a sensitivity analysis on the governing parameters, assuming they have independent effects. The results are shown in \autoref{fig:shrrate}. All but the initial gradient setup ($t_0$) show some degree of influence. By dimensional analysis the scalings are combined into the following proposed model:

\begin{equation}
    \label{shr_2}
    \frac{\partial s}{\partial t}\bigg|_{shrink}  = - c D s\left(\frac{1}{h}\right)^2,
\end{equation}
 where $c$ is a dimensionless proportionality constant. 

\begin{figure}[htp]
  \centering
    \includegraphics[width=0.88\textwidth]{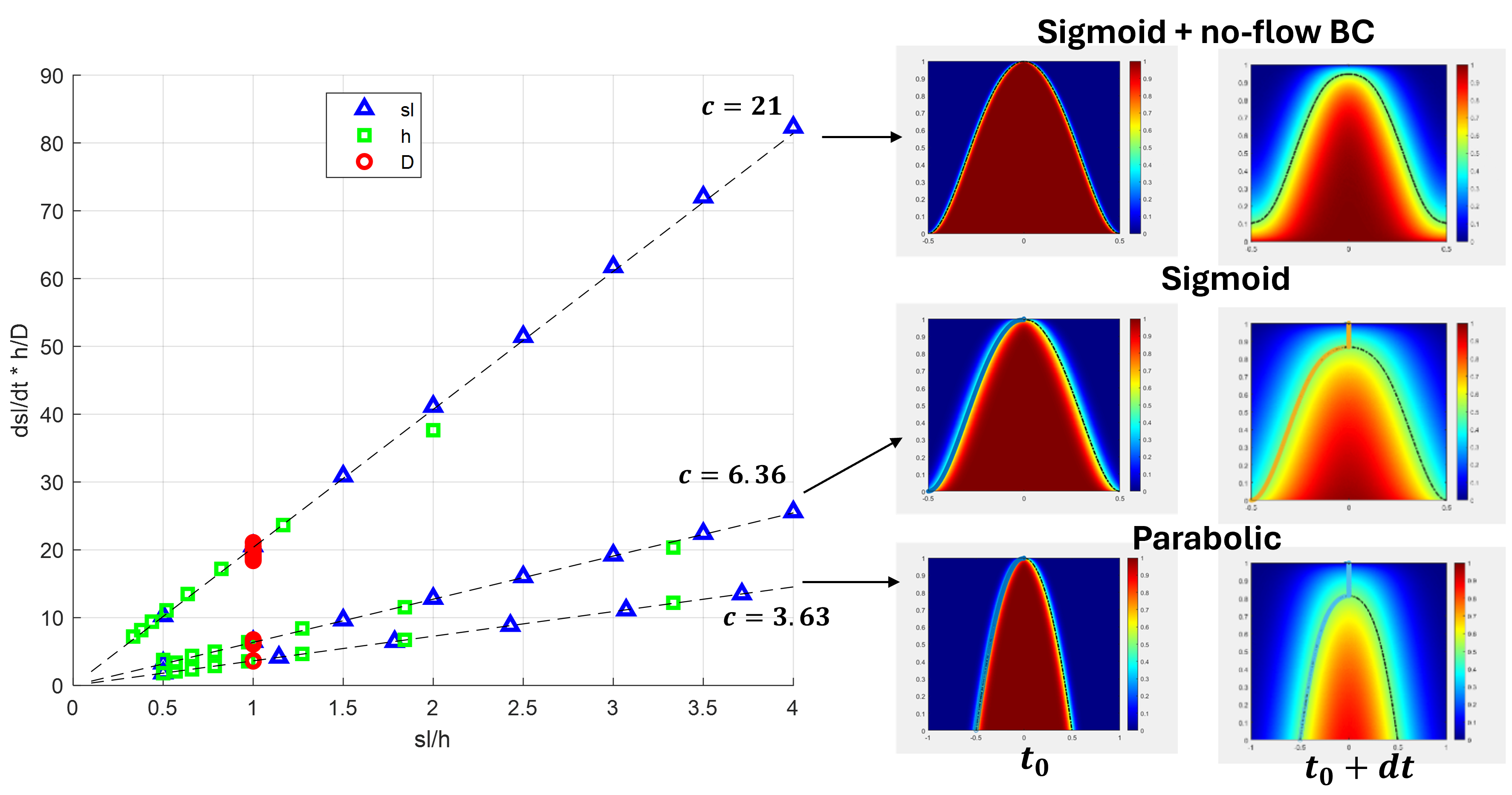}
    \caption{Empirically calculating the proportionality constant $c$ for different lamella shape setups. The shrinking rate model is found to be robust to a variety of scenarios.}
    \label{fig:generalized_splm}
\end{figure}

 The relationship presented in \autoref{shr_2} 
 is validated for various initial geometric concentration fields (e.g., sigmoidal and circular shapes), as well as alternative lateral boundary conditions such as no-flux (\autoref{fig:generalized_splm}). The sole effect all these variations have pertains to a change in the proportionality constant $c$, which provides a degree of freedom that may reflect differing flow conditions. Similarly, it has been verified (not shown here for brevity) that the same shrinking rate expression holds for a three-dimensional lamella, with the change in dimensionality affecting only the value of $c$. 
 This is not central to our discussion and a more extensive examination concerning the proportionality constant $c$ is not pursued here. 

\FloatBarrier

\end{document}